\documentclass[preprint,showpacs,amsmath,amssymb,aps,prb]{revtex4}

\usepackage{graphicx}
\usepackage{bm}
\usepackage{epsfig}

\begin{document}

\title{Ac-induced thermal vortex escape in magnetic-field-embedded
long annular Josephson junctions}
\author{Niels Gr{\o}nbech-Jensen}
\affiliation{Department of Applied Science,
 University of California, Davis, California 95616.}
\author{Matteo Cirillo}
\affiliation{Department of Physics and INFM, University of Rome "Tor Vergata",
I-00133 Rome, Italy.}
\today

\begin{abstract}
We investigate theoretically the thermal escape behavior of trapped magnetic
fluxons in long annular Josephson junctions in dc magnetic fields, and
perturbed by a probing ac current. The study is motivated by recently
published experimental data that show multi-peaked escape distributions for
increasing bias current in the extreme low temperature regime when
the system is perturbed by an ac current. We demonstrate that the observed
behavior of multi-peaked escape distributions can be reproduced and
predicted in the entirely classical, thermally driven sine-Gordon model,
which is widely accepted as accurately describing the experimental system.
We interpret the observed multi-peaked distributions as being directly
induced by dynamical resonances between the applied ac perturbation and
the natural oscillation frequency of a trapped fluxon.
\end{abstract}

\maketitle

Several experiments have suggested \cite{Martinis85,Wallraff103} that
Josephson junctions, operated in their zero-voltage state and close to the
critical current, will reveal signatures of quantum levels if perturbed by
ac currents at frequencies which, when multiplied with Planck's constant,
correspond to the energy between the anticipated levels. The signatures are
found by slowly increasing the dc bias current of the junction and recording
the current at which the voltage across the junction switches from zero to a
non-zero value. By repeating such experiments, one can obtain a distribution
of critical switching currents, which reflects aspects of the nature of the
switching device. These studies have been inspired by extensively conducted
work over the past
three decades confirming the consistency between Kramers theory for \textit{%
thermal} escape with the predicted behavior of superconducting devices at
non-zero temperature for various kinds of Josephson junction systems
\cite{Fulton74,Castellano96,Lukens,Wallraff00,Castellano00,Jensen03}.
Given the consistency in the classical, thermal regime of Josephson
junctions, similar escape measurements in the low temperature regime, where
quantum uncertainty should dominate thermal fluctuations, have been
conducted. These have
provided the possibility for non-thermal features in the measurable
switching distribution. A specific aim has been to identify signatures of
quantum levels expected in the extreme low temperature regime of the
zero-voltage junction. Such energy levels could manifest themselves through
the escape statistics if more than one energy level is statistically
possible, since a junction in an excited energy state will have a higher
probability of switching at lower bias currents than a junction in the
ground state. Thus, if multiple quantum levels can be populated, then the
switching current distribution may reflect this as a multi-peaked profile,
each peak identifying a unique quantum level. Experimental efforts have
identified and reported such multi-peaked switching distributions in a
variety of Josephson junction systems with application of a small ac
current. The rationale for this ac perturbation is that it provides the
energy (at a particular frequency) to populate specific excited energy levels.

A particularly noticeable set of observations was reported for an annular
long Josephson junction in a static magnetic field, and with a single
trapped magnetic flux quantum \cite{Wallraff203}. This system is known \cite
{Jensen91,Jensen92} to exhibit behavior similar to that of a small-area
Josephson
junction, and thus, should behave equivalently to small area junctions when
conducting switching measurements \cite{Wallraff00}. Since a magnetic
fluxon is a macroscopic
object, anomalous switching behavior related to the fluxon dynamics would
imply that macroscopic quantum dynamics had been identified. It is the aim
of this paper to investigate what can be expected from an equivalent purely
classical model of the system, and compare the results to the published
experimental results to assess the interpretation of the experiments. Thus,
the physical object of our investigations is an annular Josephson
junction of inner radius $R$ and width $W$ in the limit that the length of
the junction $2\pi(R+W/2)$ is much larger than the Josephson penetration depth 
\cite{Barone82} . The total area of the junction is $(\pi W^2+2\pi RW)$ and
therefore, if the width is much less than the inner radius, a good
approximation for the area of the annular junction is $2\pi RW$. We
make this last assumption, but we also assume that the width of the
junction is smaller than the Josephson penetration depth, meaning that we
restrict our interest to one-dimensional physics along the barrier forming
the junction. In this limit, relevant electrical parameters are the
capacitance ($\mathcal{C}$), inductance ($\mathcal{L}$), and the sub-gap
conductance ($\mathcal{G}$) per unit length of the
junction. For the case of the annular junction that we
consider, $\mathcal{C}=\varepsilon _0\varepsilon _rW/{\cal T}$ and $\mathcal{%
L}=\mu _0D/W$. Here, $\varepsilon _0$ and $\mu _0$ are, respectively, the
vacuum permittivity and permeability and $\varepsilon _r$ is the
relative dielectric constant of the oxide of the junction. The
parameter ${\cal T}$ represents the thickness of the oxide barrier while the
parameter $D={\cal T}+\lambda _1+\lambda _2$ is the magnetic thickness
of the junction. These parameters are all linked in Josephson junction
through the electromagnetic wave velocity
$1/\sqrt{{\cal C}{\cal L}}=\sqrt{({\cal T}/{\cal D})/\mu_0\varepsilon_0
\varepsilon_r}$ in the oxide barrier.

\begin{figure}[tbp]
\centerline{\epsfxsize=5.0in \epsfbox{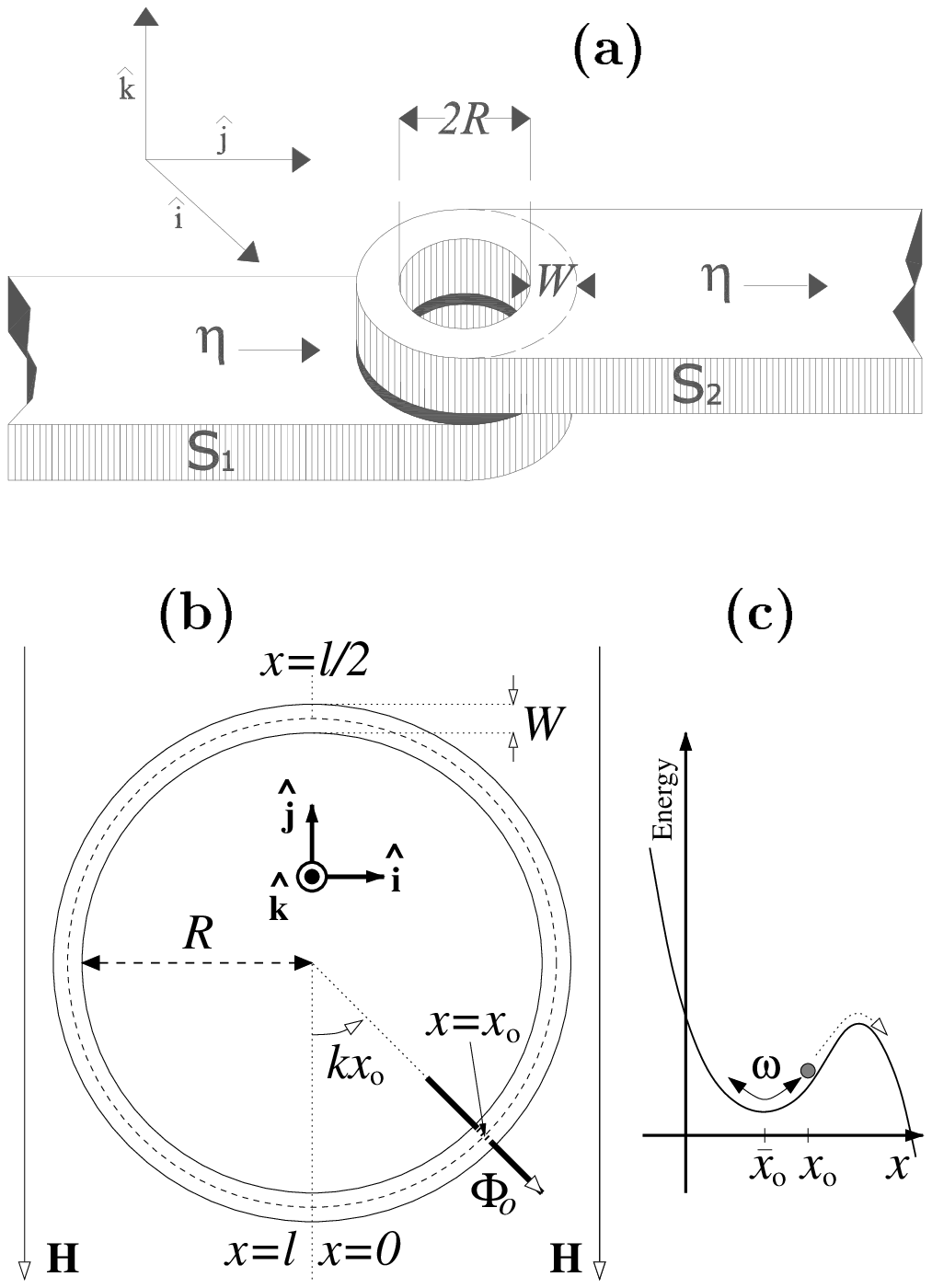}}
\caption{ (a) Sketch of the system under consideration: an annular Josephson
junction, defined by the two superconductors {\bf S}$_1$ and {\bf S}$_2$,
 with (normalized) bias $\eta=\eta_{dc}+\eta_{ac}\sin{\omega_d t}$;
(b) sketch of the trapped flux-quantum orientation and position
relative to the external magnetic field ${\bf H}$; (c)
sketch of the magnetic-field-induced washboard potential felt by the flux
quantum, which here is illustrated as a particle at position $x=x_0$.}
\label{fig:fig1}
\end{figure}

We study the perturbed sine-Gordon model in the following form, 
\begin{eqnarray}
\varphi _{tt}-\varphi _{xx}+\sin \varphi  &=&\eta _{dc}+
\eta _{ac}\sin {\omega_d t}%
+\Gamma k\sin {kx}-\alpha \varphi _t+n(x,t)  \label{eq:Eq_1} \\
\langle n(x,t)\rangle  &=&0  \label{eq:Eq_2} \\
\langle n(x,t)n(x^{\prime },t^{\prime })\rangle  &=&2\alpha \theta \delta
(x-x^{\prime })\delta (t-t^{\prime })  \label{eq:Eq_3} \\
\varphi (x,t) &=&\varphi (x+l,t)+2\pi \; ,  \label{eq:Eq_4}
\end{eqnarray}
where $\varphi (x,t)$ is the phase difference between the quantum
mechanical wave functions of the two superconductors defining the junction.
Equations (\ref{eq:Eq_1})-(\ref{eq:Eq_4}) model an annular
Josephson junction of inner
radius $R$ and width $W$ when a dc magnetic field is applied in the plane of
the junction and the forcing terms include dc and ac currents as well
as thermal
noise.
Assuming uniform Josephson supercurrent pair density $j_c$ over the area
of the junction, the critical current density (per unit length) is $Wj_c$.
This is
also the characteristic current density against which all current densities
are measured in the normalized equations (\ref{eq:Eq_1})-(\ref{eq:Eq_4}).
Tunneling of Cooper pairs of electrons is described by the $\sin \varphi$
term
length. Dissipative quasi-particle current density is represented by the term
proportional to $\alpha =\mathcal{G}\hbar \omega _p/2eWj_c$ where $\omega _p$
is the unperturbed plasma frequency $\omega _p=\sqrt{\frac{2\pi Wj_c}{\Phi _0%
\mathcal{C}}}$ and where $\Phi_0=\frac{h}{2e}\approx2.07\cdot10^{-15}Wb$ is
the flux quantum. The spatial coordinate $x$ is normalized to the Josephson
penetration depth $\lambda _J=\sqrt{\frac{\Phi _0}{2\pi \mathcal{L}%
Wj_c}}$ while the temporal coordinate $t$ has been normalized to
$\omega_p^{-1}$. An external magnetic field in the plane of the junction (see
Fig.~\ref{fig:fig1}) is represented by $\Gamma$ \cite{Jensen91,Jensen92}, which
is normalized to the characteristic field $H_0=j_c\lambda _J$.
Given the annular geometry, the effect of the
magnetic field is a spatially harmonic potential function with periodicity $%
l=\frac{2\pi (R+W/2)}{\lambda _J}$ the normalized length of the junction;
thus, $k=2\pi /l$. The junction is driven by a dc current, an ac current,
and a thermal noise current from the dissipative quasi-particle term
proportional to $\alpha $, represented by the normalized quantities $\eta
_{dc}$, $\eta _{ac}$ and $\omega_d$, as well as $n$, where $n$ is given by the
dissipation fluctuation relationship of Eqs.~(\ref{eq:Eq_2}) and (\ref
{eq:Eq_3}) \cite{Parisi88}.
The normalized temperature $\theta $ is measured relative to
the characteristic temperature $T_0=E_J/k_B$, where $E_J=\frac{\Phi _0Wj_c}{%
2\pi }$ is the Josephson energy density. Finally, since we consider only
the situation of a single
trapped flux quantum, we always impose the boundary
condition of Eq.~(\ref{eq:Eq_4}).

As outlined in Refs.~\cite{Jensen91,Jensen92}, the dynamics of an annular
Josephson
junction with a single trapped fluxon (flux quantum) should behave similarly
to a small-area Josephson junction, which in turn is typically modeled by a
pendulum equation \cite{Barone82}. This can be illustrated by defining the
wave momentum $P$,
\begin{eqnarray}
P=-\frac 18\int \varphi _x\varphi _tdx\;,
\label{eq:Eq_5}
\end{eqnarray}
and using the standard kink-soliton solution, 
\begin{eqnarray}
\varphi _s=4\tan ^{-1}\left[ \exp \left( -\frac{x-x_0-ut}{\sqrt{1-u^2}}\right)
\right] \;, \label{eq:Eq_6}
\end{eqnarray}
where $u=\dot x_0$ and the resulting fluxon momentum is $P=u/\sqrt{1-u^2}$.
We then obtain the following dynamical equation for the fluxon position, $x_0$: 
\begin{eqnarray}
\frac{dP}{dt}=\frac \pi 4\eta _{dc}+\frac \pi 4\eta _{ac}\sin {\omega_d t}%
-\alpha P-\frac \pi 4\Gamma 
k\mathrm{sech}\left( \frac 12\pi k\sqrt{1-u^2}\right)
\sin{kx_0}+\nu (t)  \; .
\label{eq:Eq_7}
\end{eqnarray}
The thermal fluctuation term associated with the fluxon is \cite{MRS}
\begin{eqnarray}
\langle \nu (t)\rangle &=&0 \label{eq:Eq_8} \\
\langle \nu (t)\nu (t^{\prime })\rangle &=&\frac 14\frac{\alpha \theta }{%
\sqrt{1-u^2}}\delta (t-t^{\prime }) \; .
\label{eq:Eq_9}
\end{eqnarray}
In the following, we restrict the analysis to the non-relativistic
limit, $\sqrt{1-u^2}\rightarrow 1$. In this limit, the above equation is
equivalent to the usual pendulum model for a small Josephson junction, and
for $\theta =\alpha =\eta _{ac}=0$, we can directly determine the critical
current $\eta _c^{(f)}$ and linear plasma oscillation of the fluxon $\omega
_p^{(f)}$: 
\begin{eqnarray}
\eta_c^{(f)} &=&\Gamma k\mathrm{sech}\left(\frac{\pi k}2\right)
\label{eq:Eq_10}
\\
\omega _p^{(f)} &=&\sqrt{\frac \pi 4k\sqrt{\left( \eta _c^{(f)}\right)
^2-\eta _{dc}^2}} \; , \label{eq:Eq_11}
\end{eqnarray}
where $\eta _c^{(f)}$ is the maximum dc bias current for which zero-voltage
states are possible in the system, and $\omega _p^{(f)}$ is the small
amplitude oscillation frequency of the fluxon in the washboard potential for 
$\eta _{dc}\le \eta _c^{(f)}$. The corresponding equilibrium location of
the static fluxon is given by 
\begin{eqnarray}
\bar{x}_0=k^{-1}\sin ^{-1}\left( \frac{\eta _{dc}}{\eta _c^{(f)}}\right) \; .
\label{eq:Eq_12}
\end{eqnarray}
These expressions are derived for $\alpha =n=\eta _{ac}=0$, $%
l\gg 1$, and small to moderate values of $\Gamma$, and arise from a collective
coordinate treatment of the fluxon. Thus, the results account neither for
spatial fluxon deformation due to the perturbation terms nor for the
periodic boundary conditions.

The very close analogy between the behavior of a long annular Josephson
junction with a single trapped fluxon in a magnetic field, and a small area
Josephson junction has led to several theoretical and experimental studies.
One of the latest is the investigation to identify macroscopic quantum
tunneling, triggered by the application of microwave radiation, of a flux
quantum through the effective washboard potential barrier \cite{Wallraff203}.
Since it was recently demonstrated \cite{Jensen04} that the signatures in
anomalous switching
distributions obtained from small-area Josephson junctions, generated by the
application of ac perturbations and attributed to quantum levels, can be
reproduced in the classical pendulum model and in experiments conducted
at temperatures well above the quantum transition temperature, it is
logical to anticipate the same overlap in signatures between the published
experimental low temperature results and the classical fluxon behavior.

To investigate the relationship between the classical model
and the experimental observations reported in Ref.~\cite{Wallraff203},
we have conducted extensive numerical simulations of the statistics of
thermal fluxon
escape from the magnetic-field-induced potential barrier in a manner similar to
the experiments. The system is initiated with a trapped fluxon, a given
magnetic field $\Gamma$, and a dc bias current well below the critical
value, $\eta _c^{(f)}$. Thermal noise and an ac bias current at a given
frequency are also applied. The dc bias current is continuously increased,
typically with a normalized rate of $\dot \eta _{dc}\le 10^{-6}$. The system
is defined to have switched from the "static" zero-voltage state by 
meeting three
conditions: 1) the fluxon position, measured as the position of the
largest value of $|\varphi _x|$, exceeds $l/2$, 2) the phase
$\langle \varphi
\rangle _x-\sin ^{-1}\eta _{dc}$ exceeds $\pi $, and 3) the normalized
voltage $\langle \dot \varphi \rangle _x$ exceeds $\frac 14$ of the
normalized voltage corresponding to the perturbation estimate of the steady
fluxon motion when $\Gamma=0$ \cite{Jensen91}: 
\begin{eqnarray}
\langle \dot \varphi \rangle _x\Big|_{\Gamma=\eta _{ac}=n=0}
=\frac{\pi \eta _{dc}}{%
\sqrt{(4\alpha )^2+(\pi \eta _{dc})^2}} \; .
\label{eq:Eq_13}
\end{eqnarray}
All three switching criteria yield nearly identical switching distributions
for $\dot \eta _{dc}\lesssim 3.5\cdot10^{-5}$. For the very fast sweeps,
one of the criteria may be activated immediately due to transient effects
from the initial conditions. In such a case, we define switching to have
happened when all three criteria are fulfilled.
The simulations are carried out with a spatial
central difference discretization of $dx=0.025$ and a Verlet temporal
integrator with time step, $dt\approx0.02$. The normalized damping
parameter has been chosen here as $\alpha=0.001$ for all data shown.

\begin{figure}[tbp]
\centerline{\epsfxsize=5.0in \epsfbox{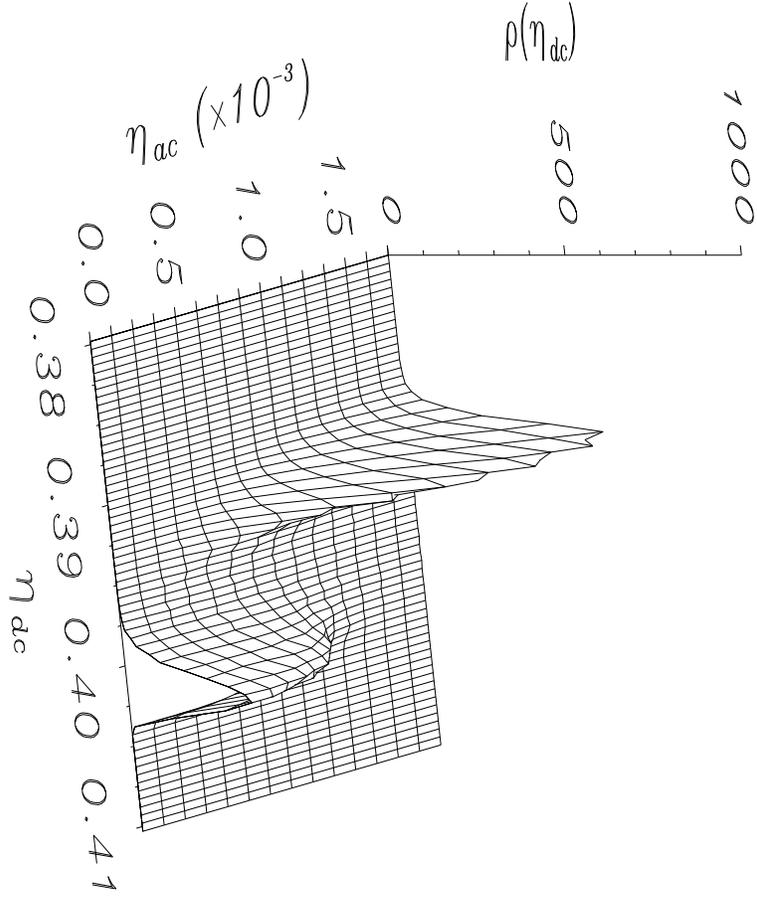}}
\caption{ Simulated switching distributions $\rho(\eta_{dc})$
for $l=10$, $\Gamma=1$, $\dot{\eta}%
_{dc}=10^{-6}$, $\omega_d=0.2$, and $\theta=2.5\cdot10^{-3}$. Each
distribution represents 5000 switching events. }
\label{fig:fig2}
\end{figure}

We first show the results of a set of simulations of 75,000 switching events
for $l=10$, $\Gamma=1$, and $\omega_d =0.2$. Figure \ref{fig:fig2} shows the
resulting
probability density $\rho (\eta _{dc})$ for switching at a given value of $%
\eta _{dc}$ for different values of the ac amplitude $\eta _{ac}$. The
density is normalized such that $\int \rho \,d\eta _{dc}=1$. For very small
values of $\eta _{ac}$, we find a single-peaked localized switching
distribution for 
$\eta _{dc}$ values just below the predicted critical switching current $%
\eta _c^{(f)}$. However, for increasing ac amplitude, we observe a
nontrivial transition to more complicated statistics of the switching
events. For $\eta _{ac}$ in the interval of $5\cdot 10^{-4}\lesssim \eta
_{ac}\lesssim 10^{-3}$, we observe clear evidence of a double-peaked
distribution, closely resembling the experimental observations of Ref.~\cite
{Wallraff203}, and for ac amplitudes higher than $10^{-3}$, we again find a
localized single-peak distribution, but centered at a lower $\eta _{dc}$
value than for $\eta _{ac}=0$. The origin of the multi-peaked distributions
observed experimentally has been interpreted as evidence for the population
of different quantum levels of the fluxon in the magnetic-field-induced
potential well, which exists in the static state. Given that the model of this
paper is classical, we must interpret our results differently.

In the classical system under investigation here, we do not have intrinsic
quantum energy levels of the macroscopic flux quantum. However, as in the
experiments, distinct
resonances between the fluxon oscillation and the ac field may appear when
the ac frequency and the fluxon plasma frequency become harmonics of one
another; specifically, when they are equal. 
The applied frequency $\omega_d=0.2$ used for the simulations of Figure
\ref{fig:fig2},
equals the fluxon resonance frequency $\omega_p^{(f)}$ in the bias current
vicinity of the secondary peak in the switching distribution for increasing
ac amplitude.

\begin{figure}[tbp]
\centerline{\epsfxsize=5.0in \epsfbox{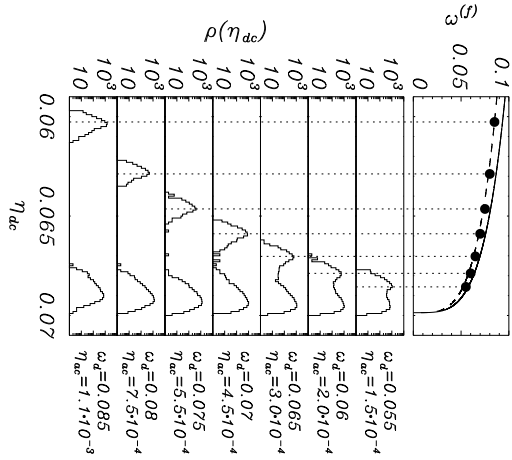}}
\caption{ Simulated switching distributions $\rho(\eta_{dc})$
for $l=20$, $\Gamma=0.25$, $\dot{\eta}%
_{dc}=2.5\cdot10^{-7}$, and $\theta=10^{-3}$. Each distribution represents
5000 switching events. Top plot shows the linear resonance (solid curve),
Eq.~(\ref{eq:Eq_11}), and the anharmonic resonance (dashed curve) of
Eqs.~(\ref{eq:Eq_18}) and (\ref{eq:Eq_21}). Markers ($\bullet$)
show the corresponding frequency-bias relationships as obtained from
the ac-induced resonant peaks of the displayed switching distributions. }
\label{fig:fig3}
\end{figure}

This is verified by Figure \ref{fig:fig3},
in which we have displayed the relationship
between the location of the secondary (the ac-induced) peak in bias current,
and the predicted fluxon resonance frequency.
To demonstrate that the multi-peak switching distribution in the classical
limit is not limited to particular parameter ranges, we have
chosen to show results
for $%
l=20$, $\Gamma=0.25$, and $\dot \eta _{dc}=2.5\cdot 10^{-7}$. These parameters
provide a critical fluxon bias current of $\eta _c^{(f)}\approx 0.0699$.
Each distribution shown in Figure \ref{fig:fig3} is for a particular value
of the driving
frequency $\omega_d $ and the corresponding ac amplitude has been optimized
to produce multiple peaks of similar content in switching events. Figure
\ref{fig:fig3}
shows the predicted, Eq.~(\ref{eq:Eq_4}), relationship between the applied
frequency and the location of the ac-induced peak as a solid curve. As can
be seen, very reasonable agreement is found. We notice, however, that the
peak location is consistently shifted slightly down in frequency (down in
current) compared to the predicted linear resonance. This slight discrepancy
is consistent for all system parameters we have tried (different lengths and
magnetic fields). The reason for the variation is that the resonance leading
to any switching event is not purely linear, since the fluxon must
experience large oscillations in at least a cubic nonlinearity in order to
escape the potential well. Thus, the resonance frequencies are depressed by
third-order nonlinearities. This can be quantified from the following simple
analysis. Let $x_0=\bar x_0+A$, where $\bar x_0$ is a constant and $A$
represents oscillatory motion of the fluxon around $\bar x_0$. The
non-relativistic, noiseless limit of Eq.~(\ref{eq:Eq_7}) then reads :
\begin{eqnarray}
\ddot A+\frac \pi 4\eta _c^{(f)}\sin {k\bar x_0}\cos {kA}+\frac \pi 4\eta
_c^{(f)}\cos {k\bar x_0}\sin {kA}=\frac \pi 4\eta _{dc}+\frac \pi 4\eta
_{ac}\sin {\omega_d t}-\alpha \dot{A} \; .
\label{eq:Eq_14}
\end{eqnarray}
We assume a single-mode solution in the form 
\begin{eqnarray}
A=a\sin (\omega_d t+\psi ) \; ,
\label{eq:Eq_15}
\end{eqnarray}
which, when inserted into Eq.~(\ref{eq:Eq_14}), yields 
\begin{eqnarray}
\sin {k\bar x_0} &=&\frac{\eta _{dc}}{\eta _c^{(f)}J_0(ka)} \label{eq:Eq_16} \\
\frac \pi 4\eta _{ac}\sin (\omega_d t-\psi ) &=&a\left[ \frac \pi 2k\eta
_c^{(f)}\frac{J_1(ka)}{ka}\cos {k\bar x_0}-\omega_d ^2\right] \sin {\omega_d t}%
+\alpha a\omega_d \cos {\omega_d t} \; . \label{eq:Eq_17} 
\end{eqnarray}
In the low dissipation limit, this results in the amplitude-dependent
resonance frequency 
\begin{eqnarray}
\omega _{res}^{(f)}=\sqrt{\frac \pi 2k\eta _c^{(f)}\frac{J_1(ka)}{ka}\sqrt{%
1-\left( \frac{\eta _{dc}}{\eta _c^{(f)}J_0(ka)}\right) ^2}} \; .
\label{eq:Eq_18}
\end{eqnarray}
Notice that for small amplitude oscillations with $|ka|\ll 1$, Equations
(\ref{eq:Eq_11}) and (\ref{eq:Eq_18}) coincide. For larger oscillation
amplitudes $a$, however, we
observe a decrease in both critical current and resonance frequency. For
multi-peaked switching distribution, as observed in Figure
\ref{fig:fig3}, obtained for
relatively low temperature $\theta $, the dynamics must be such that the
driven oscillations of the fluxon have an amplitude similar to the distance
from the energetic minimum position to the corresponding energetic saddle
position of the washboard potential; i.e., we can estimate the relevant
amplitude of oscillations for multi-peaked distributions to be given by 
\begin{eqnarray}
ka &\lesssim &\pi -2\sin ^{-1}\left( \frac{\eta _{dc}}{\eta _c^{(f)}J_0(ka)}%
\right) \label{eq:Eq_19}\\
\Rightarrow \;\;J_0(ka)\cos {\frac{ka}2} &\approx &\frac{\eta _{dc}}{\eta
_c^{(f)}} \; .
\label{eq:Eq_20}
\end{eqnarray}
A polynomial expansion of the left-hand side of Eq.~(\ref{eq:Eq_20})
for small $ka$ yields the
following approximate relationship between the fluxon oscillation amplitude
and the bias current: 
\begin{eqnarray}
ka\approx \sqrt{\frac 43\left[ 1-\frac{\eta _{dc}}{\eta _c^{(f)}}\right] } \; .
\label{eq:Eq_21}
\end{eqnarray}
Inserting Eq.~(\ref{eq:Eq_21}) into Eq.~(\ref{eq:Eq_18}) provides
an estimate of the anharmonic
resonance frequency relevant for resonant multi-peaked switching
distributions. This relationship, $\omega _{res}^{(f)}$ as a function of $%
\eta _{dc}$, is shown in Figure \ref{fig:fig3} as a dashed curve.
We clearly see the
depression of the resonance and the very good agreement with the numerical
simulations. We notice that fitting the critical bias $\eta _c^{(f)}$ in
the linear resonance curve shown in Eq.~(\ref{eq:Eq_11}) will lead to very good
agreement between simulation results as well, but it is important to
emphasize that while Eqs.~(\ref{eq:Eq_18}) and (\ref{eq:Eq_21}) are
approximate expressions they do not contain any fitting parameters.

\begin{figure}[tbp]
\centerline{\epsfxsize=5.0in \epsfbox{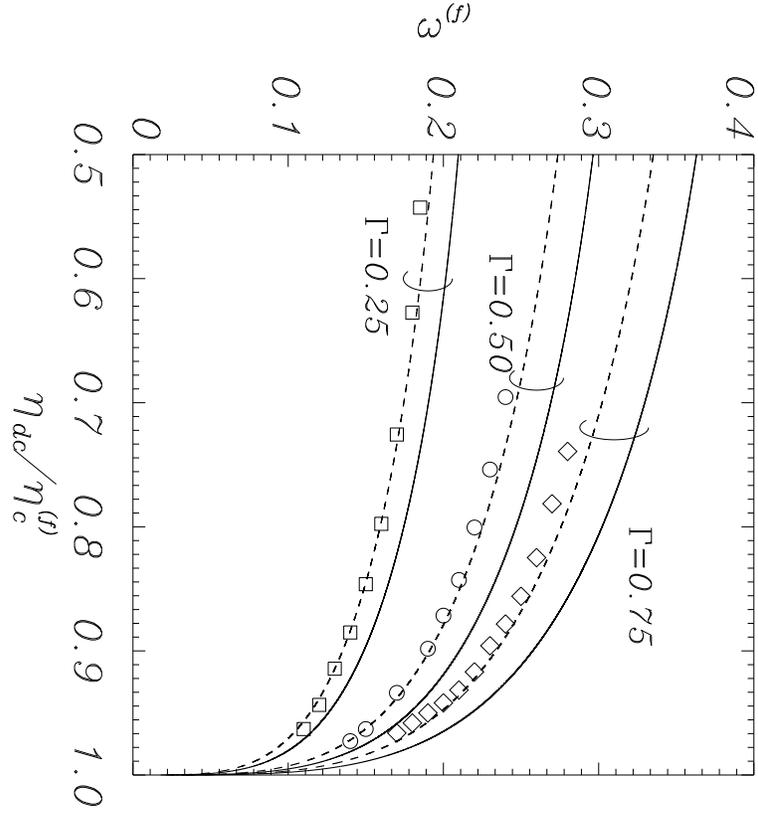}}
\caption{ Simulated relationships between bias and ac-induced oscillation
resonance for $l=10$, $\dot{\eta}_{dc}=10^{-6}$, and $\theta=2.5\cdot10^{-3}$.
Each marker ($\Gamma=0.25$ $\Box$; $\Gamma=0.5$ $\bigcirc$; $\Gamma=0.75$
$\Diamond$)
represents a distribution as shown in Figure \ref{fig:fig3} with 5000
switching events. Solid curves represent the linear resonance,
Eq.~(\ref{eq:Eq_11}), and
the dashed curves represent the anharmonic resonance of Eqs.~(\ref{eq:Eq_18})
and (\ref{eq:Eq_21}). }
\label{fig:fig4}
\end{figure}

\begin{figure}[tbp]
\centerline{\epsfxsize=5.0in \epsfbox{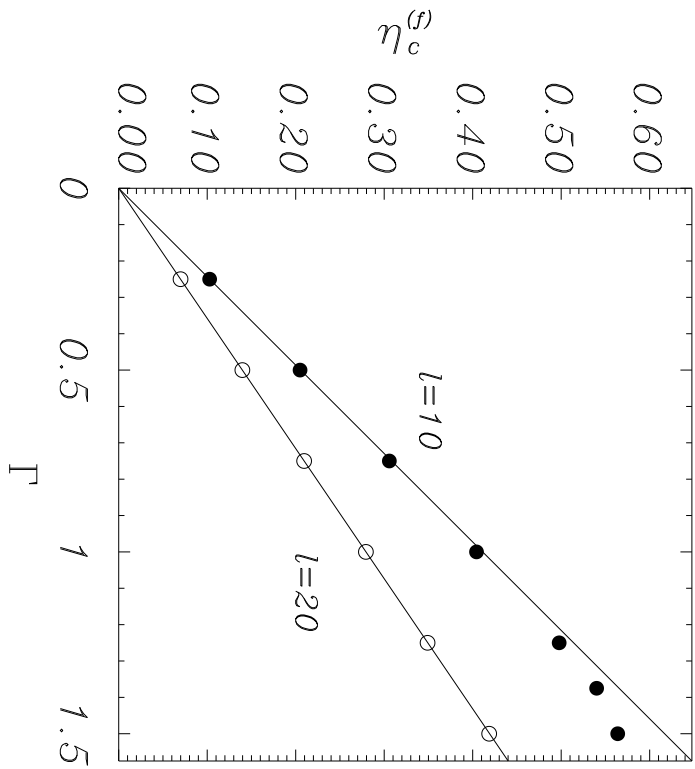}}
\caption{Comparisons between the predicted critical bias (solid lines) from
Eq.~(\ref{eq:Eq_10}) and simulated values ($l=10$ $\bullet$; $l=20$ $\bigcirc$)
for $\theta=\eta_{ac}=0$. }
\label{fig:fig5}
\end{figure}

Further investigations of the ac-induced multi-peaked switching
distributions and the relationship to the above perturbation analysis are
shown in Figure \ref{fig:fig4} for a system with parameters $l=10$, $\dot \eta
_{dc}=10^{-6}$, and $\Gamma=\frac 14,\frac 12,\frac 34$. We show here
only the locations of the ac-induced switching distribution peaks as markers
along with the corresponding linear resonance (Eq.~(\ref{eq:Eq_11}), solid) and
anharmonic resonance (Eqs.~(\ref{eq:Eq_18}) and (\ref{eq:Eq_21}), dashed).
Again we find very good
consistency between the simulated secondary resonant switching peaks and the
approximate resonance curves. We see a slight discrepancy developing for
increasing magnetic field $\Gamma$. This arises due to the strong sensitivity of
the perturbation result to the predicted critical current as given by
Eq.~(\ref{eq:Eq_10}). Figure \ref{fig:fig5} shows the predicted critical
bias as lines together with the
corresponding values obtained from numerical simulations of
Eq.~(\ref{eq:Eq_1}) for $\theta =\eta _{ac}=0$. The agreement is good
for most low to moderate
values of $\Gamma$, but larger magnetic fields show that the predicted value
overestimates the actual critical bias. However, even the very slight
difference
observed for $l=10$ and $\Gamma=\frac 34$ gives rise to the discrepancy
observed in Figure \ref{fig:fig4} for $\Gamma=\frac 34$, and a fine tuning
of $\eta _c^{(f)}$ by less
than 1\% can make the agreement in Figure \ref{fig:fig4} be near perfect.

\begin{figure}[tbp]
\centerline{\epsfxsize=5.0in \epsfbox{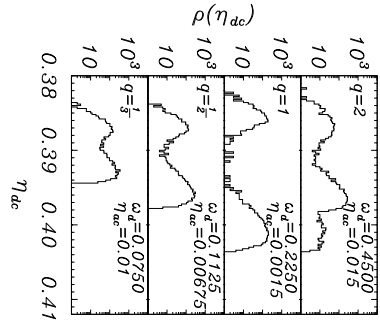}}
\caption{ Simulated ac-induced switching distributions for various harmonics
($q$) of the same frequency, $\omega_d=q\cdot0.225$. Parameters are:
$l=10$, $\Gamma=1$%
, $\dot{\eta}_{dc}=10^{-6}$, and $\theta=2.5\cdot10^{-3}$. Each distribution
represents 5000 switching events. }
\label{fig:fig6}
\end{figure}

The ac-induced resonant escape described above has been analyzed for 
situations when an
applied ac frequency equals the fluxon oscillation resonance. However, one
can equally well expect resonant escape when the applied frequency is a
simple rational multiple of the resonance frequency, thereby
exciting the
fluxon oscillations. This is demonstrated by a few switching distributions
shown in Figure \ref{fig:fig6} for parameters $l=10$ and $\Gamma=1$. Here,
$q$ indicates the
relationship between the driving frequency and the excited resonance $%
q^{-1}\omega_d =\omega _{res}^{(f)}$, giving rise to the left-most (small
bias) escape peak. As can be directly observed, both superharmonic and
subharmonic resonances can result in ac-induced switching. Furthermore, we
observe (for $q=2$) that more than one resonance may be excited enough to
provide more complex thermal switching distributions in some cases.

\begin{figure}[tbp]
\centerline{\epsfxsize=5.0in \epsfbox{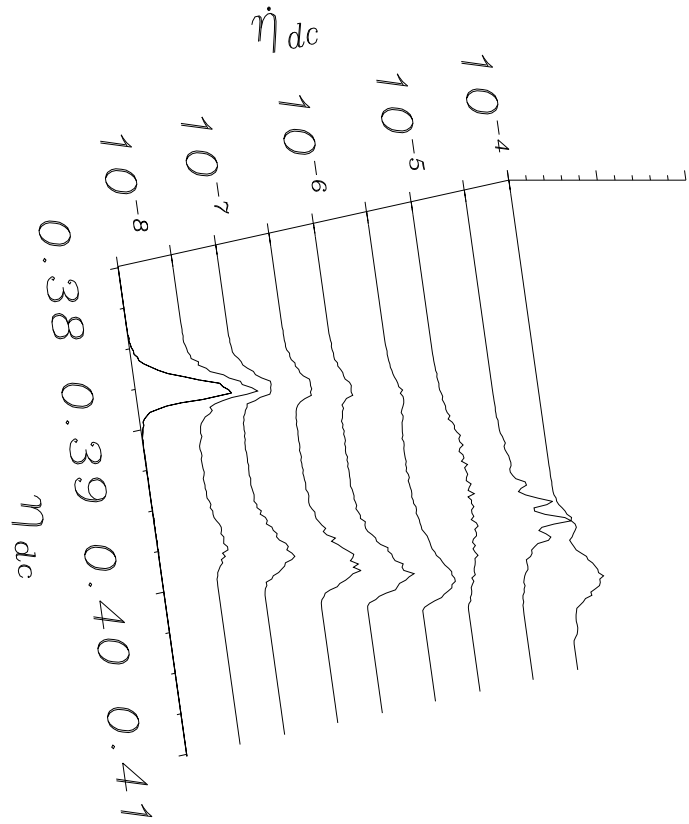}}
\caption{ Simulated switching distributions $\rho(\eta_{dc})$ for
$l=10$, $\Gamma=1$, $%
\omega_d=0.2125$, $\eta_{ac}=10^{-3}$, and $\theta=2.5\cdot10^{-3}$. Each
distribution represents at least 5000 switching events. }
\label{fig:fig7}
\end{figure}

\begin{figure}[tbp]
\centerline{\epsfxsize=5.0in \epsfbox{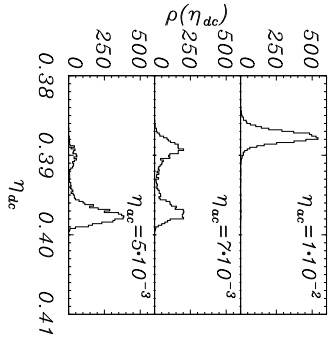}}
\caption{ Simulated switching distributions $\rho(\eta_{dc})$ for
$l=10$, $\Gamma=1$, $%
\omega_d=0.2125$, $\dot{\eta}_{dc}=10^{-8}$, and $\theta=2.5\cdot10^{-3}$
for three different ac-amplitudes: $\eta_{ac}=10^{-3}$ (top),
$\eta_{ac}=7\cdot10^{-3}$ (middle), $\eta_{ac}=5\cdot10^{-4}$ (bottom).
Each distribution represents at least 2000 switching events.
}
\label{fig:fig8}
\end{figure}

Figure \ref{fig:fig7} summarizes an investigation of the effect of the sweep
rate
$\dot \eta_{dc}$. Key parameters here are $l=10$,
$\Gamma=1$, $\omega_d =0.2125$,
and $\eta_{ac}=10^{-3}$. As should be expected, switching distributions
of different
sweep rates differ such that lower sweep rates enhance early (low $\eta
_{dc} $) switching, simply because more switching attempts can be made for
low bias values. Consequently, in the extreme slow sweep simulations ($\dot
\eta _{dc}=10^{-8}$) we observe almost all switching events near the ($q=1$)
resonance ($\eta _{dc}\approx 0.3875$). In the opposite limit ($\dot \eta
_{dc}=10^{-4}$), we observe nearly all switching events around (or
beyond) the critical bias ($\eta _{dc}\approx \eta _c^{(f)}\approx 0.41$),
and we also observe complex switching distributions for these very fast
sweeps.
In this case the sweep rate is too large for the fundamental resonance
to develop within
the time that $\eta _{dc}$ provides fluxon resonances near the driving
frequency, and the system consequently does not have significant opportunity
to develop a large amplitude resonant state. Between the two extremes of
Figure \ref{fig:fig7}, we have distributions of sweep rates that provide
distinctly
separated switching peaks. We have, therefore,
except for data shown in Figures \ref{fig:fig7} and \ref{fig:fig8},
chosen sweep rates that on the one hand provide multi-peaked distributions that
are not artifacts of excessively fast sweep rates, and on the other hand
provide simulation times that will yield statistically meaningful switching
distributions with our available computer resources. We notice that the nearly
single-peak
distribution shown for $\dot \eta _{dc}=10^{-8}$ in Figure 7
should not be interpreted
as if multi-peaked distributions vanish for slow sweep rates. In fact,
multi-peaked distributions for very slow current sweeps will become
observable if the ac amplitude $\eta _{ac}$ is slightly decreased. Likewise, the
very fast sweep will reveal the resonant behavior if the ac amplitude $%
\eta _{ac}$ is increased. Thus, we note that complex switching distributions
may be easily obtained for many meaningful parameters in ac-driven Josephson
systems. To emphasize this essential point, we have
produced switching distributions for $\dot{\eta}_{dc}=10^{-8}$ for
different values of $\eta_{ac}$. The results, shown in Figure \ref{fig:fig8},
clearly
show how lowering the ac amplitude is compensating for the low sweep rate,
thereby again illuminating the resonant switching behavior associated with
the coupling to the applied ac field.

In conclusion, we have thoroughly investigated the classical thermal
sine-Gordon analog to the recently published experimental data from an
ac-driven, magnetic-field-embedded annular Josephson junction with a single
trapped fluxon. As a result of the applied ac perturbations, we can observe
resonant behavior of the trapped fluxon, which in turn gives rise to
measurable escape signatures closely resembling experimental observations
\cite{Wallraff203} made in low temperature conditions. Simple
considerations of the (anharmonic)
resonances responsible for the anomalous switching distributions show very
good and consistent agreement between simulation and analytic perturbation
results. Since the external ac field creates resonances whose switching
distribution signatures coincide with those of possible intrinsic quantum
levels and quantum tunneling, we interpret the results as evidence that the
identification of multi-peaked switching distributions cannot be
unambiguously related to the identification of quantum levels and quantum
tunneling if the anomalous switching distribution is generated with the
application of an external ac field. This conclusion is further strengthened
by recent experimental and theoretical work \cite{Jensen04} demonstrating
that multi-peaked switching distributions in small-area Josephson junctions
can be induced by an ac perturbation for temperatures
well above the quantum transition temperature similarly to
what has been reported in the low temperature regime.

\acknowledgments
This work was supported in part by the Computational Nanoscience Group of
Motorola Inc. NGJ acknowledges generous hospitality during several visits
to the University of Rome "Tor Vergata". We are grateful to Prof.~Alan
Laub for carefully reading the manuscript.

{}

\newpage

\end{document}